\documentclass[superscriptaddress,aps,pra,twocolumn,showpacs,nofootinbib,longbibliography,notitlepage]{revtex4-2}
\usepackage{etex}
\usepackage{amsmath,amssymb,amsthm}
\usepackage[colorlinks=true,citecolor=blue,urlcolor=blue]{hyperref}
\usepackage[pdftex]{graphicx}
\usepackage{times,txfonts}
\usepackage{braket}
\usepackage{color}
\usepackage{natbib}
\usepackage{amsmath,blkarray}
\usepackage{mathtools}
\usepackage{latexsym}
\usepackage{tabularx, booktabs}
\usepackage{graphics,epstopdf}
\usepackage{graphicx}
\usepackage{float}
\usepackage{graphicx}
\usepackage{amsfonts}
\usepackage{subcaption}
\usepackage{color,soul}

\newcommand{\be}{\begin{equation}}
	\newcommand{\ee}{\end{equation}}
\newcommand{\ba}{\begin{eqnarray}}
	\newcommand{\ea}{\end{eqnarray}}

\begin{document}
	\title{Sharing nonlocality in quantum network by unbounded sequential  observers }
	\author{ Shyam Sundar Mahato }
 \email{shyamsundarmahato59@gmail.com}
 \affiliation{National Institute of Technology Patna, Ashok Rajpath, Patna, Bihar 800005, India}
 \author{ A. K. Pan }
	
	\email{akp@phy.iith.ac.in}
 \affiliation{National Institute of Technology Patna, Ashok Rajpath, Patna, Bihar 800005, India}
 \affiliation{Department of Physics, Indian Institute of Technology Hyderabad, Telengana-502284, India }

	\begin{abstract}
Of late, there has been an upsurge of interest in studying the sequential sharing of various forms of quantum correlations, viz., nonlocality, preparation contextuality, coherence, and entanglement. In this work, we explore the sequential sharing of nonlocality in a quantum network. We first consider the simplest case of the two-input bilocality scenario that features two independent sources and three parties, including two edge parties and a central party. We demonstrate that in the symmetric case when the sharing is considered for both the edge parties, the nonlocality can be shared by at most two sequential observers per edge party. However,  in the asymmetric case, when the sharing across one edge party is considered, we show that at most, six sequential observers can share the nonlocality in the network. We extend our investigation to the two-input $n$-local scenario in the star-network configuration that features an arbitrary $n$ number of edge parties and one central party. In the asymmetric case, we demonstrate that the network nonlocality can be shared by an unbounded number of sequential observers across one edge party for a suitably large value of $n$. Further, we generalize our study for an arbitrary $m$ input $n$-local scenario in the star-network configuration. We show that even for an arbitrary $m$ input scenario, the nonlocality can be shared by an unbounded number of sequential observers. However, increasing the input $m$, one has to employ more number of edge parties $n$ than that of the two-input case to demonstrate the sharing of an unbounded number of sequential observers.   

	\end{abstract}
	\pacs{} 
	\maketitle
	\section{Introduction}
	\label{SecI}
	Quantum correlations showcase their supremacy over classical resources in a  variety of information processing tasks. The most discussed quantum correlation is the one that enables the violation of the Bell inequalities \cite{bell}, a feature widely referred to as quantum nonlocality. Such a correlation cannot be replicated by any local hidden variable theory. Over the decades, the non-local correlation has found a plethora of applications in many different branches including quantum key distribution \cite{{Acin2006},{Acin2007},{powloski}}, witnessing Hilbert-space dimension \cite{{whener2008},{Gallego2010},{Ahrens2012},{Brunner2013},{Bowles2014},{pan}}, random number generation \cite{{gallego1},{colbeck},{pironio},{Acin16},{pan21},Curchod}, and many more. 
	
	Standard bipartite Bell experiment involves two spatially separated parties, Alice and Bob, who receive inputs $x \in \{ 0,1 \}$ and $y \in \{ 0,1 \}$, and return outputs $a\in \{ 0,1 \}$ and $b \in \{ 0,1\}$, respectively. In a local hidden variable model, it is assumed that the outcome of a measurement is pre-determined by a suitable hidden variable $\lambda$, and such an outcome is independent of the outcomes and measurement settings of a distant observer. The joint probability distribution of outcomes can be written in the factorized form as 
	\begin{equation}
		P(ab|xy)=\int \mu (\lambda) P(a|x,\lambda) P(b|y,\lambda) d\lambda
	\end{equation} 
	where $\mu(\lambda)$ represents the probability distribution over the hidden variable $\lambda$. It is widely known that such a factorized form of the joint probability distribution is incompatible with quantum theory. This feature is commonly demonstrated through the quantum violation of a suitable Bell inequality. For this, Alice and Bob have to share an entangled state and choose a suitable set of locally non-commuting observables.
	
	Standard multipartite Bell experiments are straightforward generalizations of the bipartite experiment where multiple parties share a common entangled state.  
	In contrast, Branciard \emph{et al.} \cite{branciard 2012} introduced a nontrivial tripartite Bell experiment in a network, commonly referred to as the bilocality scenario that features two independent sources. Network Bell experiment provides a conceptually different notion of nonlocality which is commonly demonstrated through the quantum violation of suitably formulated nonlinear bilocality inequality \cite{branciard 2010,branciard 2012}. It is also shown that all entangled pure quantum states violate bilocality inequalities \cite{Gisin 2017}. 
	
Network Bell experiment has been extensively studied in various topologies \cite{{Alejandro 2019},{Cyril Branciard 2012},{Renou 2019 },{Denis 2019},{Ming 2018},{Tavakoli 2016},{rafael},{Tavakoli 2020},{Ivan 2020},{Nicolas 2017},{Tamas 2020},{Benjamin2021},{Xavier2021},{Patricia2021},{Bancal2021}, tavakoli 2014, Tavakoli 2017,Andreoli 2017,kundu,munshi2021,munshi2022,Cavalcanti10,Tavakoli22}. Several theoretical proposals have also been  experimentally verified \cite{{Davide 2020},{Dylan 2017},{Francesco Andreoli 2017},{Gonzalo 2017}}.
	A straightforward generalization of the bilocality scenario is the $n$-locality scenario  \cite{tavakoli 2014} which features $n$ number of independent sources. In \cite{munshi2021,munshi2022}, the $n$-locality inequalities for arbitrary input scenario and their optimal quantum violations have been studied without assuming the dimension of the system. Characterization of network nonlocality and its correspondence with the bipartite Bell nonlocality has been studied \cite{Andreoli 2017, kundu, munshi2022}. Recently, genuine network nonlocality has also been introduced that cannot be traced back to Bell nonlocality \cite{tavakoli2021, Renou2022, supic2022self,pozas2022}. Self-testing protocols using the quantum network have recently been proposed \cite{supic2022, supic2022self}. Using a quantum network, it has been established \cite{renoureal,li} that the real quantum theory can be experimentally falsified, i.e.,  quantum theory inevitably needs complex numbers.
	
Sharing of various form of quantum correlations has recently been attracting increasing attention \cite{silva2014,Colbeck2020,saptarshi,Zhang 2021,Debarshi,Shashank,Akshata, Shounak,asmita,Karthik,sumit,Cheng2021,Mao2022,miklin}. The present work aims to explore the sharing of nonlocality by multiple independent observers in a quantum network. Silva \emph{et al.} \cite{silva2014} first raised the question of whether the Bell nonlocality can be shared among multiple observers acting individually and sequentially. Based on Clauser-Horne-Shimony-Holt (CHSH) inequality, they demonstrated that, at most, two independent observers on one side can sequentially share the nonlocality. Later it is shown that nonlocality can be shared for an arbitrary number of sequential independent observers \cite{Colbeck2020,Zhang 2021}. The sharing of other forms of quantum correlation viz., preparation contextuality \cite{asmita}, steering \cite{Shashank,Akshata}, coherence \cite{Shounak}, and entanglement \cite{saptarshi} has also been studied in a number of subsequent research papers. Certification of multiple unsharpness parameters \cite{Karthik,sumit,miklin} has also been demonstrated through the sequential sharing of quantum advantage in communication games. It has also been shown \cite{Cheng2021} that in a CHSH scenario, two qubits cannot both be recycled to generate Bell nonlocality between multiple independent observers on both sides. Recently, the recycling of nonlocal resources in a quantum network has been briefly studied in \cite{Mao2022,Cheng22}. The authors in \cite {Mao2022} demonstrated that in a star-network involving three edge parties, the nonlocality cannot be shared by the secondary parties, i.e., the sharing is limited to the first sequence of edge parties. Our work goes beyond their result.
		
	This work demonstrates the sequential sharing of nonlocality in a quantum network through the quantum violation of $n$-locality inequality in star-network configuration. Such a network feature $n$ independent sources, $n+1$ parties includes $n$ edge parties and a central party. We first consider the simplest network, the bilocality scenario ($n=2$) for two inputs ($m=2$) for each party. We show that in the symmetric case when the sharing is considered for both the edge parties, the nonlocality in the bilocal network can be shared by at most, two sequential observers per edge party. However, in the asymmetric case, when the sharing across only one edge party is considered, we demonstrate that at most, six sequential observers can share nonlocality. 
	
	We extend our study to the two-input $n$-local scenario in star-network. We demonstrate that in the symmetric case, the number of sequential observers in each edge remains restricted to two observers for any arbitrary $n$. Importantly, in the asymmetric case, we show that an unbounded number of sequential observers across one edge can share nonlocality for a suitably large value $n$. We note an important point that throughout this work, to derive the quantum violation of $n$-locality inequality, the dimension of the quantum system remains unspecified. Further, we generalize the sharing of nonlocality for an arbitrary $m$ input scenario in the star-network configuration. We show that even for an arbitrarily large value of $m$, in the asymmetric case, the nonlocality can be shared by an unbounded number of sequential observers across one edge party. However, in such a case, increasing the input $m$, one has to suitably increase the number of edge parties $n$  to demonstrate the sharing of nonlocality by an unbounded number of sequential observers.
	
	This paper is organized as follows. In Sec.\ref{secII}, we summarize the essence of the simplest star-network scenario i.e., the bilocality scenario ($n=2$). In Sec.\ref{secIII}, we provide the optimal quantum violation of bilocality inequality using an elegant sum-of-square (SOS) approach without assuming the dimension of the system. The sharing of nonlocality in a bilocal quantum network by multiple sequential edge observers is demonstrated for two different cases in Sec.\ref{secIV}. In the symmetric case, both edge parties perform the unsharp measurement and relay the post-measurement state to the next sequential edge observers for further measurement. In the asymmetric case, only one edge party performs the unsharp measurement, and the rest of the parties perform sharp measurements. The treatment is extended to an arbitrary $n$ edge parties in star-network for the two-input scenario in Sec.\ref{secV}, where we demonstrate the sharing of network nonlocality by an unbounded number of sequential edge observers across one edge. We extend the sharing of nonlocality in the quantum network for an arbitrary $m$ input scenario in Sec.\ref{secVI} and provide an analytical argument to demonstrate the sharing by an unbounded number of sequential observers. We summarize our results in Sec.\ref{secVII}.
	
	\section{Preliminaries: The bilocal network}
	\label{secII}
	Consider a standard three-party  Bell experiment involving three spatially separated parties, Alice$^{1}$, Bob and Alice$^{2}$, receiving inputs $x,y,z \in \{ 0,1 \}$, and  producing  outputs $a,b,c\in \{+1,-1\}$ respectively. All three parties share a common physical system in a standard Bell experiment. In a local model, they then share a common hidden variable $\lambda$ that fixes the outcomes of the measurements. The tripartite joint probability can then be written in a factorized form as 
	\begin{equation}
		\label{lcn}
		P(a,b,c|x,y,z)=\int \mu (\lambda) P(a|x,\lambda) P(b|y,\lambda) P(c|z,\lambda) d\lambda
	\end{equation}
	where $\mu(\lambda)$ is a probability distribution over $\lambda$.
\begin{figure}[ht]
		\centering
		\includegraphics[width=1.0\linewidth]{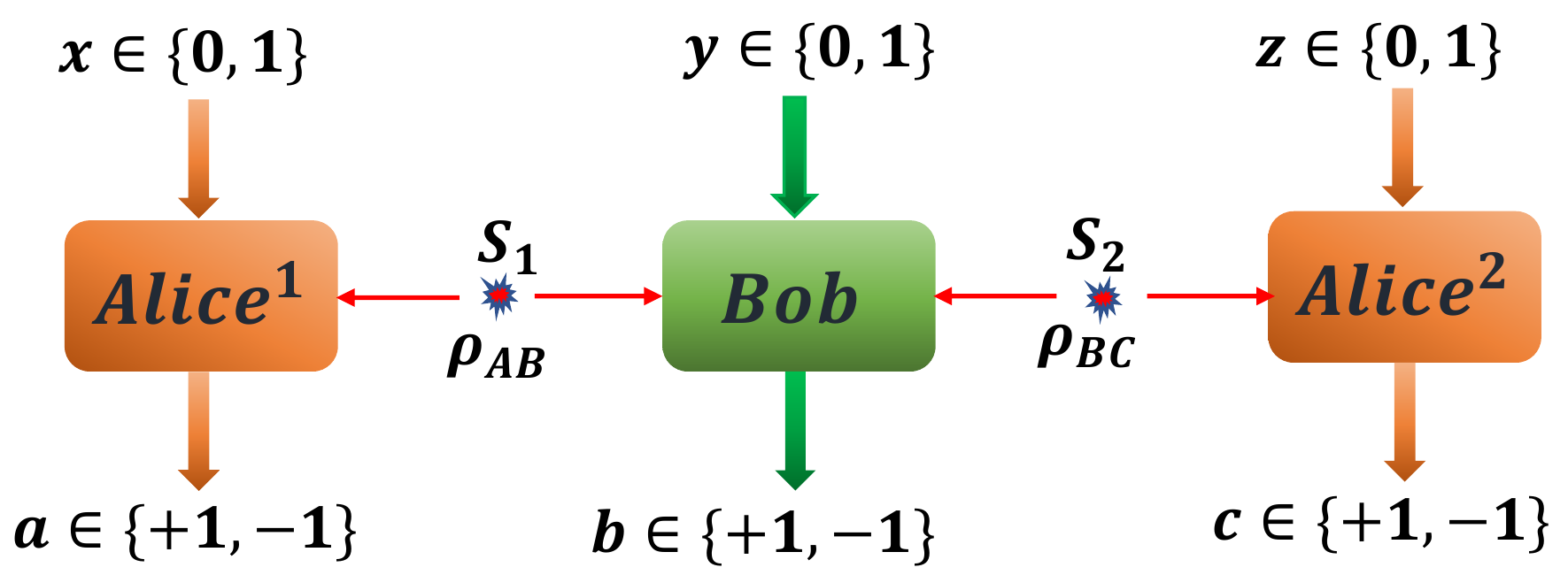}
		\caption{The bilocal scenario featuring two edge parties (Alice$^{1}$ and Alice$^{2}$) and the central party Bob. The source $S_{1}(S_{2})$ emits physical system for Alice$^{1}$ (Alice$^{2}$) and Bob. The sources are assumed to be independent to each other. In quantum theory, one may assume that the sources emit entangled states.}
		\label{bilocal}
\end{figure}
	
In contrast to the standard Bell scenario, the bilocality scenario  features two independent sources, as in Fig.\ref{bilocal}. The source $S_{1}$ ($S_{2}$) produces the physical systems $\lambda_{1}$ ($\lambda_{2}$) shared by Bob and Alice$^{1}$ (Alice$^{2}$). The crucial assumption here is that the  sources $S_{1}$ and $S_{2}$ are  independent to each other. Since the two sources to be independent the joint probability distribution $\mu (\lambda_{1}, \lambda_{2})$ can be written in  the factorized form as  $\mu (\lambda_{1}, \lambda_{2})= \mu (\lambda_{1})  \mu    (\lambda_{2})$ with $\mu (\lambda_{1})$ and $\mu (\lambda_{2})$ such that $ \int \mu (\lambda_{1}) d\lambda_{1} = \int \mu (\lambda_{2}) d\lambda_{2}=1$. This constitute the bilocality assumption in a tripartite network. Assuming bilocality the joint probability distribution can be written as
\begin{eqnarray}
\nonumber
P(a,b,c|x,y,z)=\int \int &&\mu (\lambda_{1}, \lambda_{2})P(a|x,\lambda_{1}) P(b|y,\lambda_{1},\lambda_{2})\\
&\times& P(c|z,\lambda_{2}) d\lambda_{1} d\lambda_{2}
\end{eqnarray}

The tripartite correlations in terms of the joint probability can be defined as
	\begin{equation}
		\langle A_{x}B_{y}C_{z}	\rangle =\sum_{a,b,c}^{} (-1)^{a+b+c}P(a,b,c|x,y,z)
	\end{equation}
where $P(a,b,c|x,y,z)$ is the joint probability for two-output measurements performed by Alice$^{1}$, Alice$^{2}$ and Bob. Branciard \emph{et al.} \cite{branciard 2010} proposed the following non-linear inequality
	\begin{equation}
		\label{bl22}
		\mathcal{S}_{2} \equiv \sqrt{|I|}+ \sqrt{|J|}\leq 2
	\end{equation} 
	where $I$  and $J$ are suitably defined linear combinations of these correlations are given by
	\begin{eqnarray}
		\label{ij22}
		I=\sum_{x,z=0,1}^{}\langle A_{x}B_{0}C_{z}	\rangle; \  
		J=\sum_{x,z=0,1}^{}(-1)^{x+z}\langle A_{x}B_{1}C_{z}	\rangle
	\end{eqnarray}
	
		In quantum theory, assuming that each source produces two-qubit entangled state, the optimal quantum value of $(\mathcal{S}_{2})^{opt}_{Q}=2\sqrt{2}>\mathcal{S}_{2}$.  Optimal quantum value is obtained when   Alice$^{1}$'s and Alice$^{2}$'s observables are mutually anticommuting and Bob's observables to be mutually commuting. Such choices of observables can be found for qubit system as
\begin{eqnarray}\label{observables}
A_{0}&=&C_{0}=\left(\sigma_{z}+\sigma_{x}\right)/\sqrt{2} \ \ ; \ \ B_{0}=\sigma_{z}\otimes\sigma_{z} \nonumber\\
 A_{1}&=&C_{1}=\left(\sigma_{z}-\sigma_{x}\right)/\sqrt{2} \ \ ; \ \ B_{1}=\sigma_{x}\otimes\sigma_{x}. 
\end{eqnarray}
We note here that the optimal quantum value of $(\mathcal{S}_{2})^{opt}_{Q}$ is commonly derived by assuming the dimension of the system to be a two-qubit entangled state. Here we provide a dimension-independent derivation of $(\mathcal{S}_{2})^{opt}_{Q}$. 
	
	\section{Optimal quantum bound of bilocality functional}
	\label{secIII}
To provide the dimension-independent optimization of $(\mathcal{S}_{2})_{Q}$ we introduce an elegant SOS approach which fixes the required state and the observables. We prove that there exist a suitable positive operator $\gamma$ so that $\langle \gamma \rangle=\beta- (\mathcal{S}_{2})_{Q}$, where $\beta$ is the optimal quantum value of $(\mathcal{S}_{2})_{Q}$. For our purpose, we consider that 

\begin{align}
\label{gamma1}
\langle\gamma\rangle_Q=\dfrac{\sqrt{\omega_{1}}}{2} {_{ABC}\langle}\psi|L_{1}^{\dagger} L_{1}|\psi\rangle_{ABC}+ \frac{\sqrt{\omega_{2}}}{2}{_{ABC}\langle}\psi|L_{2}^{\dagger} L_{2}|\psi\rangle_{ABC}
\end{align}
where $L_{1}|\psi\rangle_{ABC}$ and $L_{2}|\psi\rangle_{ABC}$ are suitably chosen vectors such that they satisfy the following equations. 

\begin{equation}
\label{mABC}
\begin{split}
|L_{1}|\psi\rangle_{ABC}|=\sqrt{\bigg|\left(\frac{{A}_{0}+{A}_{1}}{(\omega_{1})_{A}}
\otimes\frac{C_{0}+C_{1}}{(\omega_{1})_{C}}\right)|\psi\rangle_{ABC}
\bigg|} -\sqrt{| B_{0}|\psi\rangle_{ABC}|}\\
|L_{2}|\psi\rangle_{ABC}|=\sqrt{\bigg|\left(\frac{{A}_{0}-{A}_{1}}{(\omega_{2})_{A}}
\otimes\frac{C_{0}-C_{1}}{(\omega_{2})_{C}}\right)|\psi\rangle_{ABC}
\bigg|} -\sqrt{| B_{1}|\psi\rangle_{ABC}|}
\end{split}
\end{equation}
Here, $\omega_1=(\omega_{1})_{A}\cdot (\omega_{1})_{C}$ an $\omega_2=(\omega_{2})_{A}\cdot (\omega_{2})_{C}$ are positive numbers. For our purpose, we suitably define $(\omega_{1})_{A}=||({A}_{0}+{A}_{1})|\psi\rangle_{AB}||_{2}=\sqrt{2+\langle\{A_{0},A_{1}\}\rangle}$. Similarly, \hspace{3pt} $(\omega_{1})_{C}=\sqrt{2+\langle\{C_{0},C_{1}\}\rangle}$ \hspace{2pt} and              $(\omega_{2})_{A}=\sqrt{2-\langle\{A_{0},A_{1}\}\rangle}$, \hspace{3pt} $(\omega_{2})_{C}=\sqrt{2-\langle\{C_{0},C_{1}\}\rangle}$.

 Note that, by construction, $\langle\gamma\rangle_Q$ is positive semi-definite quantity. Now, by putting $|L_{1}|\psi\rangle_{ABC}|$ and $|L_{2}|\psi\rangle_{ABC}|$ from Eq. (\ref{mABC}) in Eq. (\ref{gamma1}), we get
\begin{align}
    \langle\gamma\rangle_{Q}=-(\mathcal{S}_{2})_{Q}         +\left(\sqrt{\omega_{1}}+\sqrt{\omega_{2}}\right)
\end{align} 	
The optimal value of $(\mathcal{S})_{Q}$ is obtained if  $\langle \gamma\rangle_{Q}=0$. This in turn provides,
\begin{eqnarray}
\label{SQopt}
\nonumber(\mathcal{S}_{2})_{Q}^{opt}&=& max[\sqrt{\omega_{1}}+\sqrt{\omega_{2}}]\\
&=&max[\sqrt{(\omega_{1})_{A}\cdot(\omega_{1})_{C}}+\sqrt{(\omega_{2})_{A}\cdot(\omega_{2})_{C} }]
\end{eqnarray}
 We use the  inequality, 
$\sqrt{r_{1}s_{1}}+\sqrt{r_{2}s_{2}}\leq\sqrt{r_{1}+r_{2}}\sqrt{s_{1}+s_{2}}$ for $r_{1}, s_{1}, r_{2}, s_{2}\geq 0$. Equality holds when $r_{1}=s_{1}$ and $r_{2}=s_{2}$. Using the inequality we can write Eq.(\ref{SQopt}) as
\begin{eqnarray}
\label{dd}
\nonumber
(\mathcal{S}_{2})^{opt}_{Q}&\leq&  \left(\sqrt{(\omega_1)_{A}+(\omega_2)_{A}}\sqrt{(\omega_1)_{C}+(\omega_2)_{C}}\right)
\end{eqnarray}
Putting the values of $(\omega_{1})_{A}$, $(\omega_{1})_{C}$, $(\omega_{2})_{A}$ and $(\omega_{2})_{C}$ it is straightforward to obtain 
\begin{align}
    (\mathcal{S}_{2})^{opt}_{Q}=2\sqrt{2}
\end{align}
when $\{A_{0}, A_{1}\}=0$ and $\{C_{0}, C_{1}\}=0$, i.e., observables of Alice$^{1}$'s as well Alice$^{2}$'s are mutually anticommuting. 
The condition $\langle \gamma_S\rangle_{Q}=0$ yields 
\begin{align}
\label{mm}
L_{1}|\psi\rangle_{ABC}=0;\hspace{2mm} 	L_{2}|\psi\rangle_{ABC}=0
\end{align} 
which provides the Bob's observables $B_{0}$ and $B_{1}$  and the state required for optimization. It is simple to check that $|\psi\rangle_{ABC}=|\psi\rangle_{AB}\otimes |\psi\rangle_{BC}$ with $|\psi\rangle_{AB}$ and $|\psi\rangle_{BC}$ are maximally entangled states in arbitrary dimension. The example in qubit system is given in Eq.(\ref{observables}). In contrast to earlier works, our derivation of $(\mathcal{S}_{2})_{Q}^{opt}$ is independent of assuming the dimension of the quantum system. 
\begin{figure*}[ht]
\centering
\includegraphics[width=1.0\linewidth]{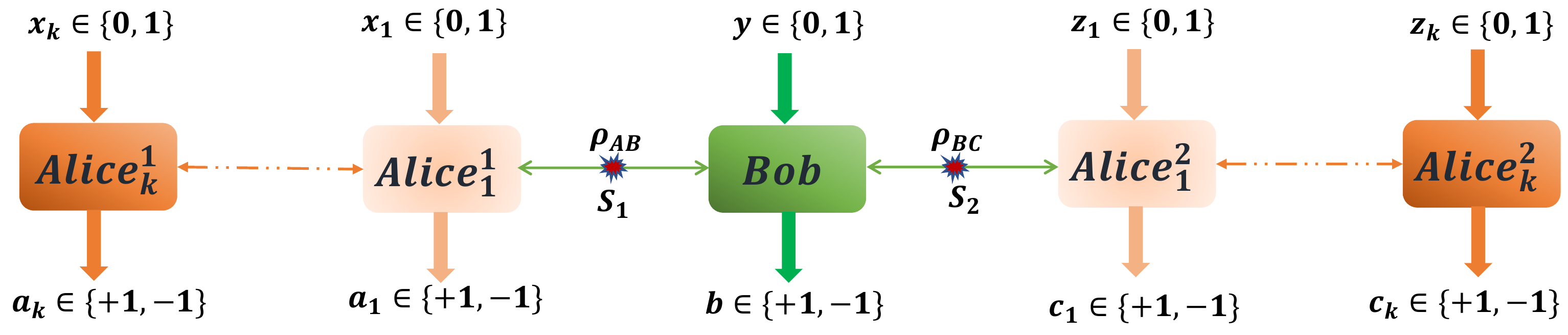}
\caption{Sequential sharing of the nonlocality in the bilocal network is depicted. There is one Bob who always performs sharp measurements and $k$ number of sequential Alice$^{1}$ and Alice$^{2}$.  Each Alice$^{1}_{k}$ performs an unsharp measurement and relays the state to the next sequential observer. A similar step is adopted by Alice$^{2}_{k}$. Here $k$ is arbitrary and the task is to determine up to which value of $k$ the nonlocality can be shared in the bilocal network.  }
\label{fig2}
\end{figure*} 

\section{Sharing of nonlocality in the bilocal network}
	\label{secIV}
	We first demonstrate the sequential sharing of nonlocality for the two edge parties in the bilocal network as depicted in Fig.\ref{fig2}. Sequential sharing of quantum correlations requires the observers to perform the unsharp measurement. Since a sharp measurement disturbs the state maximally, there remains no residual coherence or entanglement for future measurements to use. On the other hand, in an unsharp measurement the system is partially disturbed. In that case, there remains residual coherence or entanglement in the post-measurement state which can be used further. Hence, although the sharp measurement extracts full information, it is not useful for the sequential sharing of quantum correlations by sequential observers.  
	Therefore, a key requirement for sharing the quantum correlation is that the prior observers have to perform unsharp measurements which can be represented by positive operator-valued measures (POVMs). We consider the POVMs that are the noisy variant of projective measurements.
	
	In this work, we consider two different cases of sharing; the symmetric case when the nonlocality in bilocal network is considered for both the edge parties, and the asymmetric case where we explore the sharing of nonlocality for either party.
	\subsection{The symmetric case of sharing nonlocality}
	As depicted in Fig.\ref{fig2}, multiple independent Alice$^{1}$s (say, Alice$^{1}_k$) and Alice$^{2}$s (say, Alice$^{2}_k$) perform the unsharp measurements in sequence.  Here $k$ is arbitrary, and our task is to determine up to which value of $k$ nonlocal correction can be shared. We consider that there is only one Bob who always performs sharp measurements. In quantum theory, the first Alice$^{1}$ (say, Alice$^{1}_{1}$), upon receiving inputs,  randomly performs unsharp measurements on her subsystem of the entangle states she shares with Bob and relays the state to the second sequential Alice$^{1}$ (Alice$^{1}_{2}$) who does the same. A similar sequential measurement procedure are adopted by Alice$^{2}$. The process continues until $k^{th}$ Alice$^{1}$ and $k^{th}$ Alice$^{2}$ ( $k$ is arbitrary) violate the bilocality  inequality in Eq. (\ref{bl22}). We demonstrate that, at most, two Alice$^{1}$ and two Alice$^{2}$ can share nonlocality in the symmetric scenario.
	
	To implement the unsharp measurements of Alice$^{1}_{k}$ and Alice$^{2}_{k}$, we consider the joint  POVMs as $E_{xz}^{ac} = E_{x}^{a} \otimes(\mathbb{I} \otimes \mathbb{I}) \otimes E_{z}^{c}$. The related Kraus operators are $ M_{xz}^{ac} = M_{x}^{a} \otimes (\mathbb{I} \otimes \mathbb{I}) \otimes M_{z}^{c}$ such that $E_{xz}^{ac}= \left(M_{xz}^{ac}\right)^{\dagger} M_{xz}^{ac} $. Here, the inputs $x,z\in\{0,1\}$ and the outputs $a,c\in \{+1,-1\}$. We also consider the unbiased POVMs characterizing the unsharp measurements of $k^{th}$ Alice as 
	\begin{eqnarray}
		E_{x/z}^{\pm}&=&\left( \dfrac{1\pm \lambda_{2,k}}{2} \Pi ^{+}_{x/z} + \dfrac{1 \mp \lambda_{2,k}}{2} \Pi ^{-}_{x/z} \right)
	\end{eqnarray}
	where $\Pi ^{\pm}_{x/z}$ are the projectors satisfying $\Pi ^{+}_{x/z}+\Pi ^{-}_{x/z}=\mathbb{I}$, with $A_{x}=\Pi ^{+}_{x}-\Pi ^{-}_{x}$ and $C_{z}=\Pi ^{+}_{z}-\Pi ^{-}_{z}$. Consequently, the measurement operators are given by
	\begin{eqnarray}
		\label{kraus}
		M_{x/z}^{\pm}&=& \sqrt{\dfrac{1\pm \lambda_{2,k}}{2}} \Pi ^{+}_{x/z} + \sqrt{\dfrac{1 \mp \lambda_{2,k}}{2}} \Pi ^{-}_{x/z}
	\end{eqnarray}	
	where $\lambda^{1}_{2,k}\in [0,1]$ is the unsharpness parameter of Alice$^{1}_{k}$. Here subscript $2$ in $\lambda^{1}_{2,k}$ denotes bilocality scenario, i.e., $n=2$. We consider the similar POVMs for $k^{th}$  Alice$^{2}_{k}$  but with a different unsharpness parameter $\lambda^{2}_{2,k}\in [0,1]$.
	
	Now, after the unsharp measurements of  first sequences of  Alice$^{1}$ and Alice$^{2}$ (i.e., Alice$^{1}_{1}$ and Alice$^{2}_{1}$) the  post-measurement state relayed to second sequences of edge observers (Alice$^{1}_{2}$ and Alice$^{2}_{2}$) is given by 
	\begin{eqnarray}
		\label{reduced}
		\rho_{ABC}^{2}= \dfrac{1}{4} \sum_{\substack{a,c=+,- \\ x,z = 0,1}}^{} \left(M_{xz}^{ac}\right)^\dagger \rho_{ABC}  \left(M_{xz}^{ac}\right)
	\end{eqnarray}
	Repeating this process for $(k-1)^{th} $ times, the final joint state shared by Alice$^{1}_{k}$, Alice$^{2}_{k}$ and Bob can be written as 
	\begin{eqnarray}
		\label{rhok}
		\rho_{ABC}^{k}= \dfrac{1}{4} \sum_{\substack{a,c=+,-\\ x,z = 0,1}}^{} \left(M_{xz}^{ac}\right)^\dagger \rho^{k-1}_{ABC}  \left(M_{xz}^{ac}\right)
	\end{eqnarray}
	For our purpose, we write the joint measurement operators by using Eq. (\ref{kraus}) in the following form   
	\begin{eqnarray}
		\label{ek}
		\nonumber
		M^{++}_{xz}&=& \left( \gamma_{{k}}^{+} \mathbb{I} + \gamma_{{k}}^{-} {A_{x}}\right)\otimes (\mathbb{I} \otimes \mathbb{I}) \otimes \left( \delta_{{k}}^{+} \mathbb{I} + \delta_{{k}}^{-} {C_{z}}\right) \\
		\nonumber
		M^{+-}_{xz}&=& \left( \gamma_{{k}}^{+} \mathbb{I} + \gamma_{{k}}^{-} {A_{x}}\right)\otimes (\mathbb{I} \otimes \mathbb{I}) \otimes \left( \delta_{{k}}^{+} \mathbb{I} - \delta_{{k}}^{-} {C_{z}}\right)\\
		M^{-+}_{xz}&=& \left( \gamma_{{k}}^{+} \mathbb{I} - \gamma_{{k}}^{-} {A_{x}}\right)\otimes (\mathbb{I} \otimes \mathbb{I}) \otimes \left( \delta_{{k}}^{+} \mathbb{I} + \delta_{{k}}^{-} {C_{z}}\right)\\
		\nonumber
		M^{--}_{xz}&=&\left( \gamma_{{k}}^{+} \mathbb{I} - \gamma_{{k}}^{-} {A_{x}}\right)\otimes (\mathbb{I} \otimes \mathbb{I}) \otimes \left( \delta_{{k}}^{+} \mathbb{I} - \delta_{{k}}^{-} {C_{z}}\right)
	\end{eqnarray}
	where,
	\begin{eqnarray}
		\nonumber
		\gamma_{{k}}^{\pm}&=&\dfrac{1}{2\sqrt{2}}\left( \sqrt{1+\lambda^{1}_{2,k}} \pm \sqrt{1-\lambda^{1}_{2,k}} \right)\\
		\delta_{{k}}^{\pm}&=&\dfrac{1}{2\sqrt{2}}\left( \sqrt{1+\lambda^{2}_{2,k}} \pm \sqrt{1-\lambda^{2}_{2,k}} \right)
	\end{eqnarray}
	Using Eq. (\ref{ek}), the state $\rho_{ABC}^{k}$ in Eq. (\ref{rhok}) can explicitly be written as 
	\begin{eqnarray}
		\label{rhok1}
		\rho_{ABC}^{k}&=&  4\left({\gamma_{k}^{+}} {\delta_{k}^{+}} \right)^2 \rho_{ABC}^{k-1} \\
		\nonumber
		&+& 2\left( {\gamma_{k}^{+}}  {\delta_{k}^{-}}\right)^2 \left( \sum_{z=1}^{2}(\mathbb{I} \otimes \mathbb{I} \otimes \mathbb{I} \otimes C_{z}) \rho_{ABC}^{k-1} (\mathbb{I} \otimes \mathbb{I} \otimes \mathbb{I} \otimes C_{z}) \right) \\
		\nonumber
		&+&2\left( {\delta_{k}^{+}} {\gamma_{k}^{-}} \right)^2\left( \sum_{x=1}^{2}  ( A_{x} \otimes \mathbb{I} \otimes \mathbb{I} \otimes \mathbb{I}) \rho_{ABC}^{k-1} ( A_{x} \otimes \mathbb{I} \otimes \mathbb{I}\otimes \mathbb{I}) \right)\\
		\nonumber
		&+& \left( \sum_{ \substack{x,z=1}}^{2} \left( {\gamma_{k}^{-}}  {\delta_{k}^{-}}\right)^2 ( A_{x} \otimes \mathbb{I} \otimes \mathbb{I} \otimes C_{z}) \rho_{ABC}^{k-1} ( A_{x} \otimes \mathbb{I} \otimes \mathbb{I} \otimes C_{z}) \right)
	\end{eqnarray}
	Using Eq. (\ref{rhok1}), the combination of correlations $I_{k}$ and $J_{k}$  in Eq.(\ref{ij22}) for $k^{th}$ Alice$^{1}$ and $k^{th}$ Alice$^{2}$ can be derived as
\begin{eqnarray}
\label{I22}
\nonumber
I_{k} =\dfrac{\lambda^{1}_{2,k} \lambda^{2}_{2,k} }{4^{k-1}} \left[ \prod_{j=1}^{k-1} \left(1+\sqrt{1-\left( \lambda^{1}_{2,j}\right)^2} \right) \left(1+\sqrt{1-\left(\lambda^{2}_{2,j} \right)^2}\right)\right] I\\
\nonumber
J_{k} =\dfrac{\lambda^{1}_{2,k} \lambda^{2}_{2,k}}{4^{k-1}} \left[ \prod_{j=1}^{k-1} \left(1+\sqrt{1-\left(\lambda^{1}_{2,j} \right)^2} \right) \left(1+\sqrt{1-\left(\lambda^{2}_{2,j} \right)^2} \right)\right] J\\
\end{eqnarray}
	Hence, the condition for violating the bilocality inequality by any $k^{th}$ sequential Alice$^{1}$(Alice$^{1}_{k}$) and Alice$^{2}$(Alice$^{2}_{k}$) is 
	\begin{equation}
		\label{Ik22}
		(\mathcal{S}_{2})^{k}_{Q}=\sqrt{|I_{k}|}+\sqrt{|J_{k}|} >2   
	\end{equation}
	Putting the values of Eq.(\ref{I22}) in Eq.(\ref{Ik22}) we obtain that for the violation of bilocality inequality for the first sequence of edge observers Alice$^{1}_{1}$ and Alice$^{2}_{1}$, the condition 
	\begin{equation}
		\label{25}
		\sqrt{\lambda^{1}_{2,1} \lambda^{2}_{2,1}} (\mathcal{S}_{2})^{opt}_{Q} > 2
	\end{equation}
	has to be satisfied. Here $\lambda^{1}_{2,1}$ and $\lambda^{2}_{2,1}$ are the unsharpness parameter for Alice$^{1}_{1}$ and Alice$^{2}_{1}$.
	
	We define critical values $\left( \lambda^{1}_{2,1}\right)^{\ast}$ and $\left( \lambda^{2}_{2,1}\right)^{\ast}$  of the unsharpness parameters  which are just enough to get the violation of bilocality inequality. The values of unsharpness parameters less than $\left( \lambda^{1}_{2,k}\right)^{\ast}$ and $\left( \lambda^{2}_{2,k}\right)^{\ast}$ do not provide the quantum violation of bilocality inequality.  Hence, from Eq.(\ref{25}), the critical values of unsharpness parameters of Alice$^{1}_{1}$ and Alice$^{2}_{1}$  can be obtained as  $\left( \lambda^{1}_{2,1}\right)^{\ast}= \left( \lambda^{2}_{2,1}\right)^{\ast}=1/\sqrt{2}$ respectively. Using those critical values we can estimate the upper bound on $\lambda^{1}_{2,2}$ and $\lambda^{2}_{2,2}$ for Alice$^{1}_{2}$ and Alice$^{2}_{2}$, respectively. Then, for  sharing the nonlocality for second sequence of edge observers  Alice$^{1}_{2}$ and Alice$^{2}_{2}$, the following condition 
\begin{equation}
\left(\frac{\lambda^{1}_{2,2} \lambda^{2}_{2,2}}{2} \left[ \left(1+\sqrt{1- \left( \left( \lambda^{1}_{2,1}\right)^{\ast}\right)^2} \right) \left(1+\sqrt{1- \left( \left( \lambda^{2}_{2,1}\right)^{\ast}\right)^2} \right)\right]\right)^{1/2}> 1
\end{equation}
has to be satisfied. By considering the same unsharpness parameter for each edge party, we found that for the violation of bilocality by Alice$^{1}_{2}$ and Alice$^{2}_{2}$  the value of unsharpness parameters need to be $\lambda^{1}_{2,2}= \lambda^{2}_{2,2} \geq 0.8284$. Following the above procedures, we find that  for violation of bilocality inequality by the third sequences of edge parties i.e., for Alice$^{1}_{3}$ and Alice$^{2}_{3}$, the following condition  

\begin{eqnarray}\label{3rdseq}
&&\Bigg[\frac{\lambda^{1}_{2,3} \lambda^{2}_{2,3}}{8}\Bigg\{\left(1+\sqrt{1- \left( \left( \lambda^{1}_{2,1}\right)^{\ast}\right)^2} \right) \left(1+\sqrt{1- \left( \left( \lambda^{2}_{2,1}\right)^{\ast}\right)^2} \right)\nonumber\\
&& \times\left(1+\sqrt{1- \left( \left( \lambda^{1}_{2,2}\right)^{\ast}\right)^2} \right)\left(1+\sqrt{1- \left( \left( \lambda^{2}_{2,2}\right)^{\ast}\right)^2} \right)\Bigg\}\Bigg]>1
\end{eqnarray}

has to satisfied. Now, using the critical values of unsharpness parameters of first sequence observer Alice$^{1}_{1}$ and Alice$^{2}_{1}$ and the second sequence observers Alice$^{1}_{2}$ and Alice$^{2}_{2}$, the unsharpness parameter we obtained for third sequence of observers is $\lambda^{1}_{2,3}= \lambda^{2}_{2,3} \geq 1.0619$, which are not the legitimate values.
Thus, in the symmetric scenario of the sharing at most two sequential Alice$^{1}$ and Alice$^{2}$  can share nonlocality in bilocal network. However, this scenario becomes different if we consider the asymmetric case of sharing.

\subsection{The asymmetric case of sharing nonlocality}
Next, we examine whether it is possible to share nonlocality in the bilocal network by more than two sequential edge observers. We demonstrate that it is indeed possible in the asymmetric case of sharing when only one edge party considers the sharing the nonlocality. In this case, we consider that there is only one Bob and one Alice$^{2}$  who perform sharp measurements. There is an arbitrary $k$ number of sequential Alice$^{1}$s who perform unsharp measurements. In that case, from Eq.(\ref{I22}), the combination of correlations defined in Eq. (\ref{ij22}) can be written for $k^{th}$ Alice$^{1}$ as 
	\begin{eqnarray}
		\label{Ias22}
		I_{k} =\dfrac{\lambda_{2,k}}{2^{k-1}} \left[ \prod_{j=1}^{k-1} \left(1+\sqrt{1-\lambda_{2,j}^2} \right)\right] I\\
		\nonumber
		J_{k} =\dfrac{\lambda_{2,k}}{2^{k-1}} \left[ \prod_{j=1}^{k-1} \left(1+\sqrt{1-\lambda_{2,j}^2} \right)\right] J\\
		\nonumber
	\end{eqnarray}
	Here we drop the superscript 1 in $\lambda_{2,k}$ as there is only one edge party Alice$^{1}_{1}$ is sharing the nonlocality. The quantum violation of bilocality inequality is obtained when the condition   
	\begin{eqnarray}
		\left( \dfrac{\lambda_{2,k}}{2^{k-1}} \left[ \prod_{j=1}^{k-1} \left(1+\sqrt{1-\lambda_{2,j}^2} \right)\right]\right)^{1/2}(\mathcal{S}_{2})^{opt}_{Q}>2
	\end{eqnarray}
	is satisfied. Following the earlier argument we can find the critical values in the asymmetric case as $\lambda_{2,k}^{\ast}$ for Alice$^{1}_{k}$. The lower bound for the unsharpness parameter for Alice$^{1}_{1}$ is $\lambda_{2,1}\geq 1/2$, so that $\lambda_{2,1}^{\ast}= 1/2$.  Note that, the critical value in the asymmetric case is considerably lower than the value $1/\sqrt{2}$ obtained in the symmetric case. Hence, there is a possibility of sharing  the nonlocality by higher number of sequential observers than that of the symmetric case.  We found that at most six sequential  Alice$^{1}$ can share nonlocality. The bounds on unsharpness parameter $\lambda_{2,k}$ required for violating bilocality inequality in Eq. (\ref{bl22}) are given by $\lambda_{2,1}>0.50$, $\lambda_{2,2}>0.53$, $\lambda_{2,3}>0.58$, $\lambda_{2,4}>0.64$, $\lambda_{2,5}>0.72$, and $\lambda_{2,6}>0.85$.
	The critical value of unsharpness parameter $\lambda_{2,7}$ for  Alice$^{1}_{7}$ is found
	to be $\lambda_{2,7}^{\ast}=1.13$ which is not a legitimate value. 
	
	One may then ask whether more sequential observers across one edge can share nonlocality in a network for $n$-local scenario where $n$ may be taken to be arbitrarily large. Based on the sequential quantum violation of $n$-locality inequality, we show that in the asymmetric case  an unbounded number of sequential observers across one edge can share nonlocality in the star-network for suitable value of $n$.
	
	\section{Sharing of nonlocality in the $n$-local star-network for two-input scenario}
	\label{secV}
	\begin{figure*}[ht]
		\centering
		\includegraphics[width=0.9\linewidth]{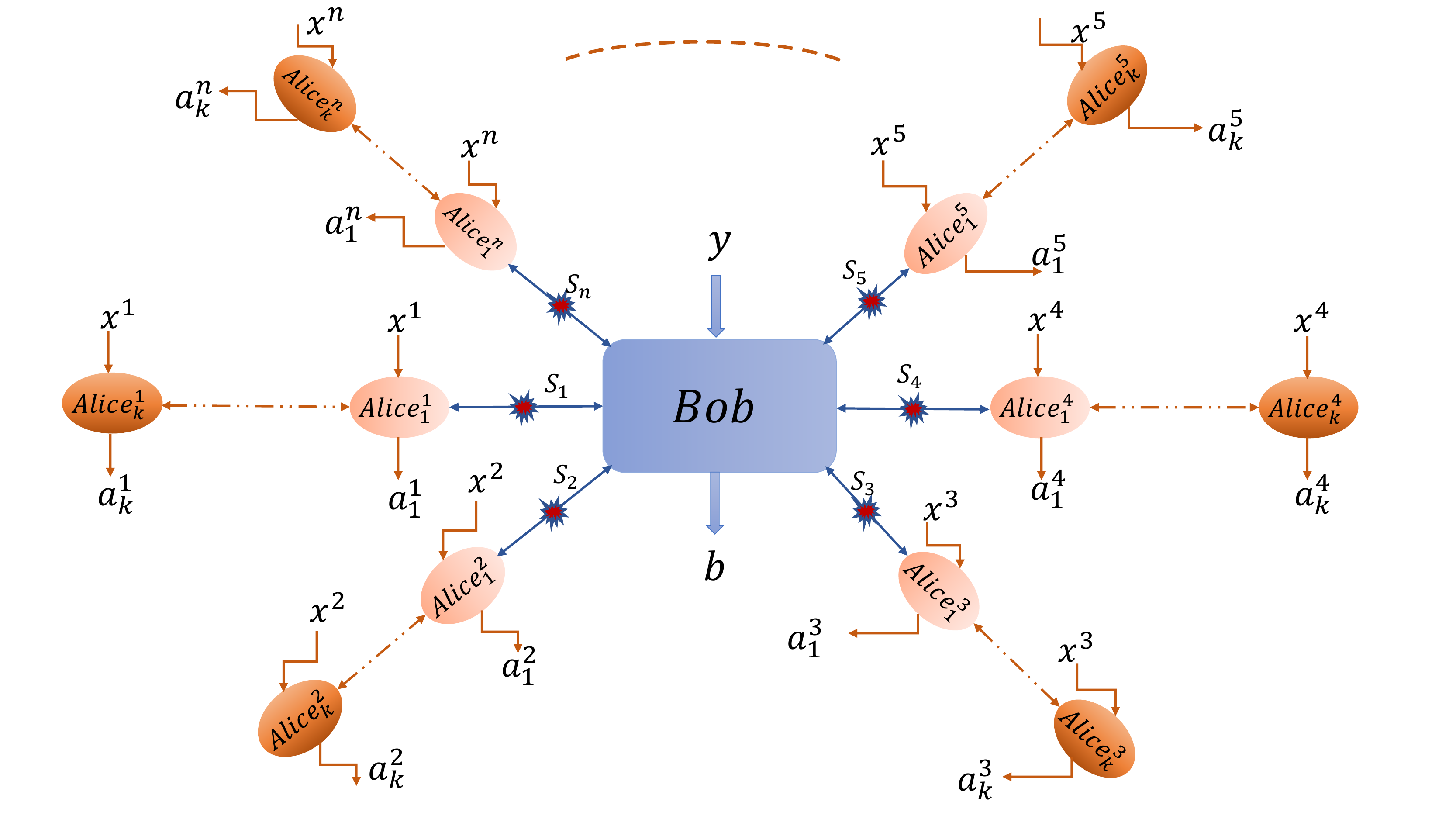}
		\caption{The $m$-input $n$-local scenario in star-network configuration consisting $n$ edge parties (Alice$^{n}$). Each edge party shares a physical system with the central party, Bob. It is assumed that sources $S_{n}$ are independent to each other. Each Alice receives $m$  inputs, i.e.,  $x^n \in [m]$ and Bob receives $y\in \left[2^{m-1}\right]$. Each edge  party Alice$_{k}^{n}$ performs unsharp measurements and relay the  state to the next sequential edge observer. }
		\label{nmlocality}
	\end{figure*}
	Let us consider a star-network configuration that features an arbitrary $n$ independent sources and  $n + 1$ parties including a central party (referred to as Bob) and  $n$ number of edge parties  (referred to as Alice$^{n}$). Each source emits a physical system to one edge party and the central party, i.e., Bob possesses $n$ number of physical systems. We also consider that each party receives only two inputs and produces two outputs. In this scenario, the $n$-locality inequality can be written as \cite{tavakoli 2014} 
	\begin{equation}
		\label{sn}
		\mathcal{S}_{n}=|I|^{1/n}+|J|^{1/n} \leq 2   
	\end{equation}
	
	The optimal quantum value is  $(\mathcal{S}_{n})^{opt}_{Q}=2\sqrt{2}$, which can also be derived in a dimension-independent manner by using our approach provided earlier.

To investigate the sharing of nonlocality in a star-network scenario, we consider that there is only one Bob who always performs sharp measurements. In the symmetric case, each of the $n$ Alices (Alice$^n$) performs unsharp measurements sequentially. For example, the edge party Alice$^1$ ($n=1$) multiple independent Alice$^{1}$s (Alice$_{k}^{1}$ with $k$ is arbitrary)  perform the unsharp measurements in sequence. A similar argument holds for every Alice$^{n}$. Here $k$ is arbitrary, and our task is to determine up to which value of $k$ nonlocality in a star-network can be shared. We demonstrate that in the symmetric case, at most, two Alice$^{n}$  can share nonlocality in the star-network.

\subsection{Sharing of nonlocality in trilocal ($n=3$) case}
We first consider the case of $n=3$, i.e.,  the trilocality scenario, to examine whether it is possible to share the nonlocality for more sequential observers compared to the bilocality scenario ($n=2$). If each of the three edge parties  performs the unsharp measurement on the respective subsystem and relays the post-measurement state to the second sequential edge observers, then the condition for getting the nonlocal advantage by any $k^{th}$ sequential Alice$^{1}$(Alice$^{1}_{k}$), Alice$^{2}$(Alice$^{2}_{k}$) and Alice$^{3}$(Alice$^{3}_{k}$) will be 

\begin{equation}
\label{Ik33}
(\mathcal{S}_{3})^{k}_{Q}=|I_{k}|^{\frac{1}{3}}+|J_{k}|^{\frac{1}{3}}>2   
\end{equation}
In trilocality scenario the expressions for $I_k$ and $J_k$ can be derived as

\begin{eqnarray}
I_{k} &=&\dfrac{\lambda^{1}_{3,k} \lambda^{2}_{3,k} \lambda^{3}_{3,k} }{8^{k-1}} \left[ \prod_{j=1}^{k-1}  \prod_{l=1}^{3} \left(1+\sqrt{1-\left( \lambda^{l}_{3,j}\right)^2} \right)\right] \  I\label{I331}\\
J_{k}&=&\dfrac{\lambda^{1}_{3,k} \lambda^{2}_{3,k} \lambda^{3}_{3,k} }{8^{k-1}} \left[ \prod_{j=1}^{k-1} \prod_{l=1}^{3}  \left(1+\sqrt{1-\left( \lambda^{l}_{3,j}\right)^2} \right)\right] \ J \label{I332}
\end{eqnarray}

Following the same procedure as adopted in symmetric bilocality scenario we have evaluated that how many sequential edge observer can share nonlocality. Plugging the expressions of Eqs. (\ref{I331}-\ref{I332}) into Eq. (\ref{Ik33}) we obtain the condition for violating the trilocality inequality by the first sequential edge observers is as follows
\begin{equation}
\label{3locality}
\left(\lambda^{1}_{3,1}\lambda^{2}_{3,1}\lambda^{3}_{3,1}\right)^{1/3} (\mathcal{S}_{3})^{opt}_{Q} > 2
\end{equation}
	
where $\lambda^{1}_{3,1} $, $\lambda^{2}_{3,1} $ and $\lambda^{3}_{3,1} $ are the unsharpness parameters for Alice$^{1}_{1}$, Alice$^{2}_{1}$ and Alice$^{3}_{1}$ respectively. 
	
From Eq.(\ref{3locality}), we find the critical values of unsharpness parameters of Alice$^{1}_{1}$, Alice$^{2}_{1}$ and Alice$^{3}_{1}$ are $\left(\lambda^{1}_{3,1} \right)^{\ast}=\left(\lambda^{2}_{3,1} \right)^{\ast}=\left(\lambda^{3}_{3,1} \right)^{\ast}=1/\sqrt{2}$. Using those critical values we can estimate the upper bound on $\lambda^{1}_{3,2} $, $\lambda^{2}_{3,2} $ and $\lambda^{3}_{3,2} $ for the second sequence of edge observers, i.e.,  Alice$^{1}_{2}$, Alice$^{2}_{2}$ and Alice$^{3}_{2}$  respectively. 
We find that for the violation of trilocality inequality (putting $n=3$ in Eq.~\ref{sn}) for Alice$^{1}_{2}$, Alice$^{2}_{2}$ and Alice$^{3}_{2}$ the values of unsharpness parameters need to be $\lambda^{1}_{3,2} = \lambda^{2}_{3,2} = \lambda^{3}_{3,2}> 0.8284$. Following the above procedures, we get that  for Alice$^{1}_{3}$, Alice$^{2}_{3}$ and Alice$^{3}_{3}$ the values of unsharpness parameters has to be $\lambda^{1}_{3,3} = \lambda^{2}_{3,3} = \lambda^{3}_{3,3}\geq 1.0619$, which are not legitimate values. Thus, in the symmetric scenario, at most two sequential observers in each edge can share the nonlocality in a star-network for $n=3$. It is straightforward to show that for any arbitrary $n$ the results remains same as $n=2, 3$, i.e., at most two sequential observers in each edge can share nonlocality in the symmetric case of sharing in star-network.
	
\begin{figure}[ht]
\centering
\includegraphics[width=1.0\linewidth]{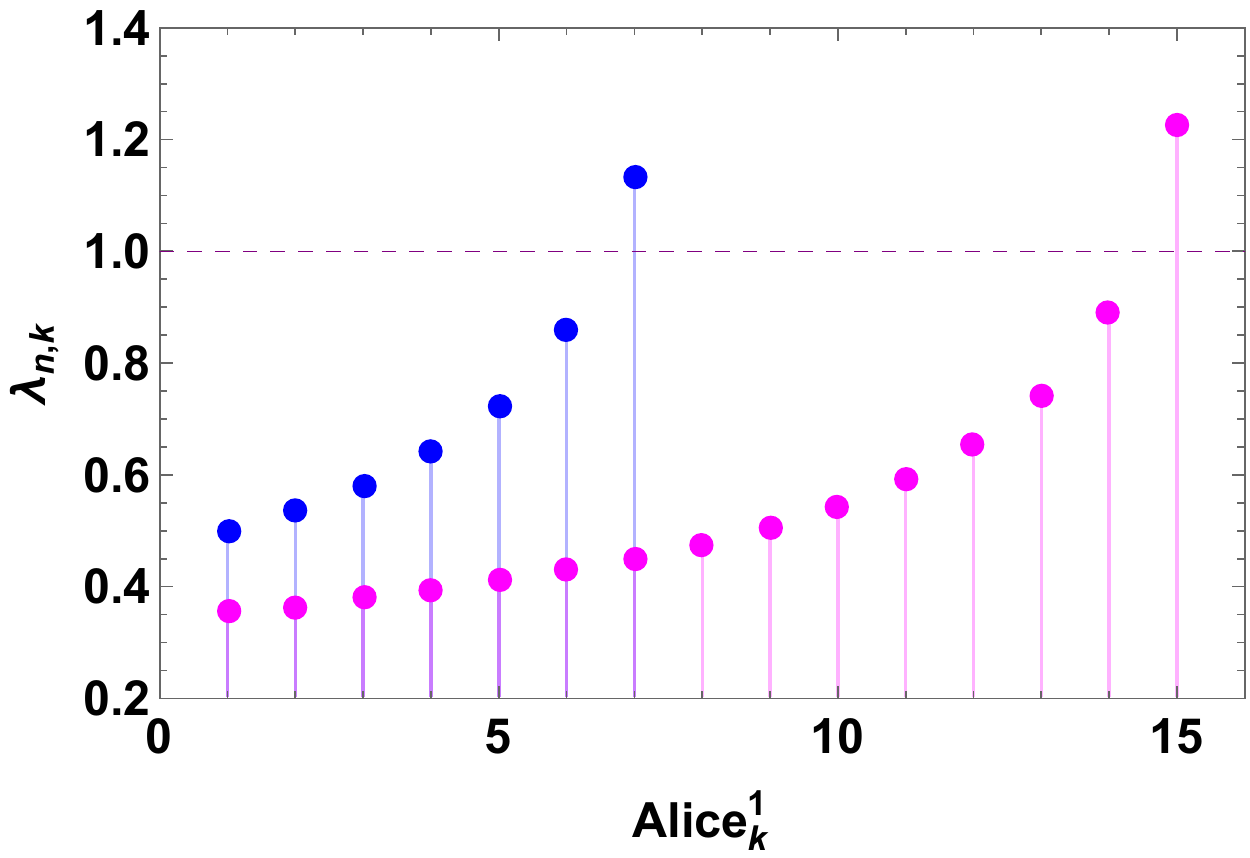}
\caption{Critical values of unsharpness parameters $\lambda_{n,k}$ required for violating the two-input ($m=2$) $n$-locality inequality are shown for $k^{th}$ observer across one edge Alice$^{1}$. The blue and magenta dots denote the critical values corresponding to the violation of bilocality ($n=2$) and trilocality ($n=3$) inequalities, respectively. For $n=2$ and $n=3$ at most six and fourteen sequential observers can share the nonlocality in quantum network.}
\label{nlocal1}
\end{figure}

	In the asymmetric case of sharing, we demonstrate that a higher number of sequential observers across one edge can share nonlocality for $n=3$ case compared to $n=2$ case (bilocality scenario). In Fig.\ref{nlocal1}, we show that at most fourteen sequential observers ($k=14$) can share nonlocality for $n=3$ when one edge observer (say, Alice$^{1}$) performs unsharp measurement sequentially while all the other edge parties perform sharp measurement.

\subsection{Sharing of nonlocality in the asymmetric $n$-local case}
Further, we generalize the above results in the asymmetric case in star-network for arbitrary $n$. It is intuitively followed that an unbounded number of sequential observers across one edge can share nonlocality in the $n$-locality scenario in the asymmetric case as $n$ is not bounded. We provide analytical proof of the above statement with legitimate approximations. For this, we consider once again that all the edge parties Alice$^{n}$ except Alice$^{1}$ perform the sharp measurement. 
The condition for obtaining the violation $n$-locality inequality in Eq. (\ref{sn}) for $k^{th}$ sequential observer across one edge (Alice$^{1}$) is given by
\begin{eqnarray}
\label{2nlocal}
\left(\dfrac{\lambda_{n,k}}{2^{k-1}} \left[ \prod_{j=1}^{k-1} \left(1+\sqrt{1-\lambda_{n,j}^2} \right)\right]\right)^{1/n}(\mathcal{S}_{n})^{opt}_{Q} > 2
\end{eqnarray}
	where $\lambda_{n,j}$ is the unsharpness parameter for $j^{th}$ sequence of Alice$^{1}$ in the $n$-locality scenario.
	
	For any arbitrary $k^{th}$ observer Alice$^{1}_{k}$, we derive the general condition on the unsharpness parameter for violating the $n$-locality inequality from Eq. (\ref{2nlocal}) as
	\begin{equation}
		\label{unscdn}
		\lambda_{n,k}\geq \dfrac{2\lambda_{n,k-1}}{1+\sqrt{1-\lambda^{2}_{n,k-1}}}
	\end{equation}
	Note that from Eq.(\ref{2nlocal}) we can obtain the critical value of unsharpness parameter for the first sequential edge observer Alice$^{1}_{1}$ which is $(\lambda_{n,1})^{\ast}= \frac{1}{2^{n/2}}$. Using this critical value of the unsharpness parameter of the $(k-1)^{th}$ sequential edge observer we observe that $1+\sqrt{1-\lambda^{2}_{n,k-1}} \geq 2\sqrt{1-\lambda^{2}_{n,k-1}}$.  Using this in Eq.(\ref{unscdn}), one can demand that the lower bound on $\lambda_{n,k}$ requires
	\begin{equation}
		\label{unsharp1}
		\lambda_{n,k}\geq \frac{\lambda_{n,k-1}}{\sqrt{1-\lambda^{2}_{n,k-1}}} 
	\end{equation}
	Eq.(\ref{unsharp1}) is obtained by using legitimate approximation and, for small $n$, it overestimates the lower bound of the unsharpness parameter. By noting the critical value of the unsharpness parameter required for the first observer  and by using Eq.(\ref{unsharp1}), we can estimate the lower bound $\lambda_{n,2}\geq 1/ \sqrt{2^{n}-1 } $ for second sequential observer Alice$^{1}_{2}$. Similarly, for the third observer  $(k = 3)$, putting the critical values of $\lambda_{n,1}$ and $\lambda_{n,2}$ in Eq.(\ref{unsharp1}) we get $\lambda_{n,3}\geq 1/ \sqrt{2^{n}-2 } $. We can then generalize the argument for any arbitrary number of $k$. Thus, if the $k^{th}$ sequential observer across one edge shares nonlocality then the following condition on unsharpness parameter
	\begin{equation}
		\lambda_{n,k}\geq \dfrac{1}{\sqrt{2^{n}-(k-1)}  }
	\end{equation}
	has to be satisfied. Alternatively, for a given $k$, one can find a $ n\equiv n(k)$ for which nonlocality in star-network can be shared by the $k^{th}$ sequential Alice$^{1}$ is given by
	\begin{equation}
		2^{n}\geq k-1+\frac{1}{\lambda^{2}_{n,k}}
	\end{equation}
	If the $k^{th}$ measurement is considered to be sharp, i.e., $\lambda_{n,k}=1$, we then have
	\begin{equation}
		\label{nk}
		n(k)\geq \log_2 k
	\end{equation}
	Thus, given an arbitrary $k$, there exists a $n(k)$ for which the nonlocality can be shared across the one edge of the star-network. Since $n$ is unbounded, we claim that an unbounded number of sequential observers can share nonlocality in the star-network in the asymmetric case of sharing. 
	
	However, there can be many different cases of sharing are possible. For example, instead of considering the sharing across one edge, one may consider sharing for two or more edges. Following our above treatment, those cases can also be dealt straightforwardly.\\
	
\section{Sharing nonlocality in star-network for arbitrary input scenario} \label{secVI}
	Let us now consider the case when each edge party Alice$^{n}$ in the star-network receives $m$ number of inputs and the central party Bob receives $2^{m-1}$ number of inputs. There are $n$ independent sources $S_{n}$ and $n+1$ parties, as depicted in Fig.\ref{nmlocality}.  To study the sharing of nonlocality, we again consider that the central party Bob always performs sharp measurements. In the symmetric case, each of the $n$ Alices (Alice$^n$) performs unsharp measurements sequentially. For example, the edge party Alice$^1$ (for $n=1$) multiple independent Alice$^{1}$s (Alice$_{k}^{1}$) performs unsharp measurements in sequence. A similar description holds for every Alice$^{n}$. Once again, our task is to determine up to which value of $k$ nonlocality in a star-network can be shared. 
	
	For this, we consider the  generalized $n$-locality inequality \cite{munshi2021} for arbitrary $m$ input  scenario is given by
	\begin{equation}
		\label{deltam}
		\mathcal{S}^{m}_{n}=\sum\limits_{i=1}^{2^{m-1}}|J^{n}_{m,i}|^{\frac{1}{n}}\leq \alpha_{m}
	\end{equation} 
	
	where \begin{equation}
		\alpha_{m}=\sum\limits_{j=0}^{\lfloor\frac{m}{2}\rfloor}\binom{m}{j}(m-2j)
	\end{equation}
	and $	J^{n}_{m,i}$s are the linear combinations of suitable correlations defined as 
	\begin{equation}
		\label{nbell}
		J^{n}_{m,i}=\bigg(\sum_{x_{1}=1}^{m}(-1)^{y_{x_{1}}^{i}}{A^{1}_{x_{1}}}\bigg) B_{i}\bigg(\sum_{x_{2}=1}^{m}(-1)^{y_{x_{2}}^{i}}{A^{2}_{x_{2}}}\bigg)	
	\end{equation}
	Here ${y_{x_{1}}^{i}}$ takes value either $0$ or $1$ and same for ${y_{x_{2}}^{i}}$.  As explained in detail \cite{munshi2021}, the values of ${y_{x_{1}}^{i}}$ and ${y_{x_{2}}^{i}}$ can be fixed by invoking the encoding scheme used in random-access-codes  \cite{Ambainis,Ghorai2018,pan,asmita} as a tool. This will fix $1$ or $-1$ values of $(-1)^{y_{x_{1}}^{i}}$ and $(-1)^{y_{x_{2}}^{i}}$ in Eq. (\ref{nbell}) for a given $i$. Let us consider a random variable $y^{\delta}\in \{0,1\}^{m}$ with $\delta\in \{1,2...2^{m}\}$. Each element of the bit string can be written as $y^{\delta}=y^{\delta}_{x_{1}=1} y^{\delta}_{x_{1}=2} y^{\delta}_{x_{1}=3} .... y^{\delta}_{x_{1}=m}$. For  example, if $y^{\delta} = 011...00$ then $y^{{\delta}}_{x_{1}=1} =0$, $y^{{\delta}}_{x_{1}=2} =1$, $y^{{\delta}}_{x_{1}=3} =1$ and so on. Now, we denote  the length $m$ binary strings as $y^{i}$ those have $0$ as the first digit in $y^{\delta}$. Clearly, we have $i\in \{1,2...2^{m-1}\}$ constituting the inputs for Bob. If $i=1$, we get all zero bit in  the string $y^{1}$ leading us $(-1)^{x_{1}^{i}}=1$ for every $x_{1}\in \{1,2 \cdots m\}$.

	In \cite{munshi2021} the optimal quantum value of $({\mathcal{S}}^{m}_{n})_{Q}$ was derived as
	\begin{align}
		\label{Copt}
		({\mathcal{S}}^{m}_{n})_{Q}^{opt}= 2^{m-1}\sqrt{m}
	\end{align}
	which is larger than $n$-locality bound in Eq. (\ref{deltam}) for  any arbitrary $m$, thereby implying the quantum violation of the $n$-locality inequality given by Eq. (\ref{deltam}).
	
	We first examine the sharing of bilocality ($n=2$)  for the symmetric case by considering three-input scenario ($m=3$). If each of the two edge parties  performs the unsharp measurement on its respective subsystem and relays the post-measurement state to the second sequential edge observers, the condition for violating the bilocality inequality for the first sequential edge parties is given by
	\begin{equation}
		\label{42}
		\sqrt{\lambda^{1}_{2,1}\lambda^{2}_{2,1}} (\mathcal{S}^{3}_{2})^{opt}_{Q} > 6
	\end{equation}
	where $\lambda^{1}_{2,1} $  and $\lambda^{2}_{2,1} $ are the unsharpness parameter for Alice$^{1}_{1}$ and Alice$^{2}_{1}$ respectively. Putting Eq.(\ref{Copt}) in Eq.(\ref{42}) for $m=3$, critical values of unsharpness parameters of Alice$^{1}_{1}$ and Alice$^{2}_{1}$ can be obtained as $\left(\lambda^{1}_{2,1}\right)^{\ast}=\left(\lambda^{2}_{2,1}\right)^{\ast}= 0.96$. Using those critical values for Alice$^{1}_{1}$ and Alice$^{2}_{1}$ we can estimate the upper bound $\left(\lambda^{1}_{2,2}\right)^{\ast}\geq 1.94$ and $\left(\lambda^{2}_{2,2}\right)^{\ast}\geq 1.94$  for Alice$^{1}_{2}$ and Alice$^{2}_{2}$ respectively, which is outside the legitimate range of the unsharpness parameter. Hence, in the symmetric case sharing of nonlocality in three-input bilocality scenario by the second sequence of edge observers is not possible. It is also straightforward to understand that at most one observer in each edge can share nonlocality in $n$-locality scenario for  $m=3$. A simple calculation can establish this fact.
	
	However, in the asymmetric case, where Bob and Alice$^{2}$ perform sharp measurements while Alice$^{1}$ continues with sequential unsharp measurements, we demonstrate that the sharing of nonlocality in bilocal network for more than one edge observer. The critical values  of the unsharpness parameters $\lambda^{\ast}_{n,k}$ required for violating bilocality inequality for $m=3$ for  the first (Alice$^{1}_{1}$) and second (Alice$^{1}_{2}$) sequential observers are given by $\lambda^{\ast}_{2,1}=0.65$ and  $\lambda^{\ast}_{2,2}=0.77$, respectively. However, the critical value of unsharpness parameter $\lambda^{\ast}_{2,3}$ for the third  sequence of Alice$^1$  (Alice$^{1}_{3}$) is $ 1.02$. Hence, the sharing of nonlocality is not possible for the third sequential observer across the edge party  Alice$^{1}$.
	\begin{figure}[ht]
		\centering
		\includegraphics[width=1.0\linewidth]{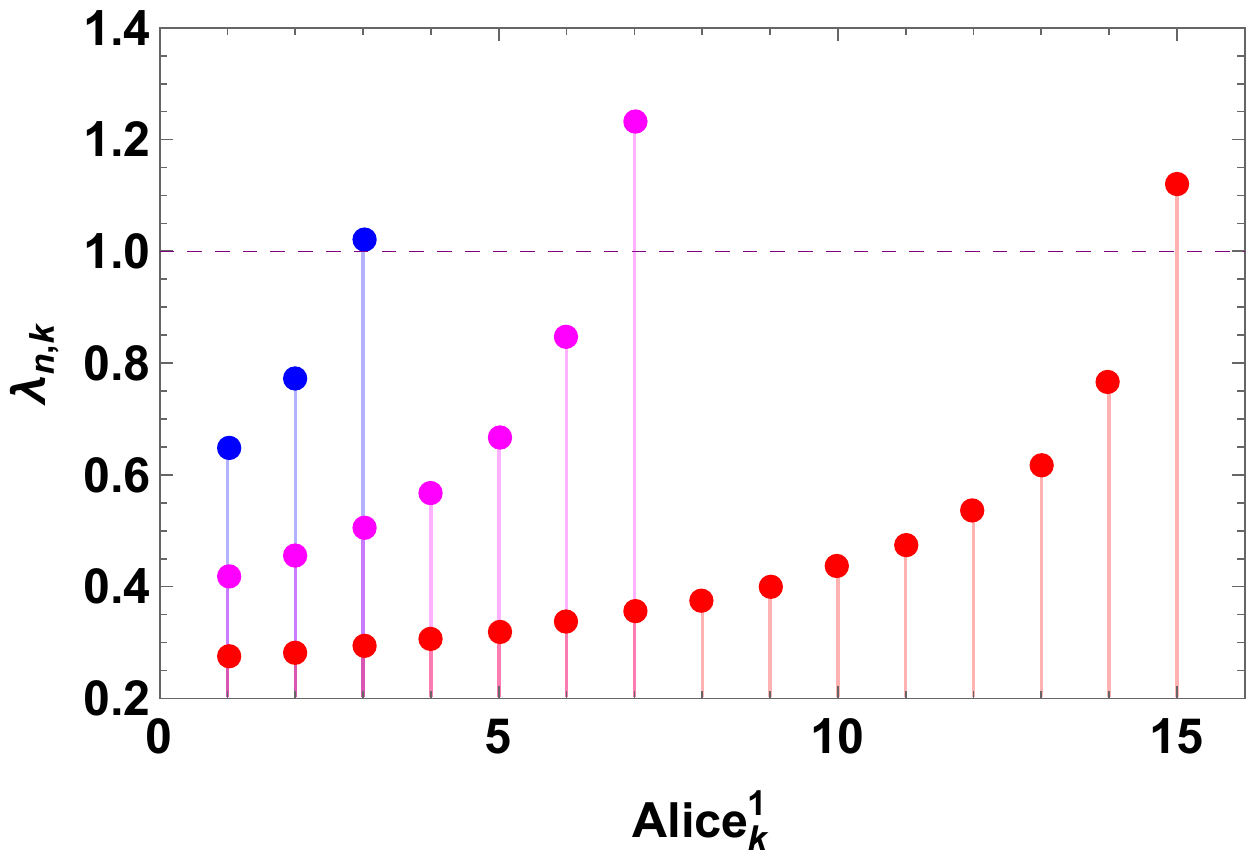}
		\caption{Critical values of unsharpness parameter $\lambda_{n,k}$ required for violating the three-input ($m=3$) $n$-locality inequality are shown for one edge party Alice$^{1}_{k}$. The blue, magenta and red dots denote the critical values corresponding to $n=3$, $n=6$ and $n=9$ respectively}
		\label{nlocal2}
	\end{figure}
	
	In Fig. \ref{nlocal2}, we  demonstrate that for three-input ($m=3$) per each edge party, the critical values of unsharpness parameter $\lambda^{\ast}_{n,k}$ required for violating the $n$-locality inequality for $k^{th}$ sequential Alice$^{1}$ (Alice$^{1}_{k}$) in the asymmetric case. The blue, magenta and red dots indicate the number of sequential Alice$^{1}$ that can share nonlocality for $n=3$, $n=6$, and $n=9$, respectively. It can be seen that by increasing the value $n$, the number of sequential observers that can share nonlocality increases. This indicates that there exists a $n$ in the star-network scenario for which the sharing of nonlocality can be demonstrated for an unbounded number of sequential observers across one edge. 
	
	We analytically prove the above statement for an arbitrary $m$ input $n$-locality scenario in star-network by using legitimate approximations. 
	By considering the asymmetric case of sharing, we find the condition for violating the arbitrary $m$ input $n$-locality inequality is given by  
	\begin{eqnarray}
		\label{nlocal}
		\left( \dfrac{\lambda^{m}_{n,k}}{m^{k-1}} \left[ \prod_{j=1}^{k-1} \left(1+\left( m-1\right)\sqrt{1-\left({\lambda^{m}_{n,j}} \right)^2} \right)\right]\right)^{1/n}(\mathcal{S}^{m}_{n})^{opt}_{Q}\geq \alpha_{m}
	\end{eqnarray}
	where $\lambda^{m}_{n,j}$ is the unsharpness parameter for $j^{th}$ sequence of Alice$^{1}$ in the $n$-locality scenario. Then, for $k^{th}$ observer, the general condition on the unsharpness parameter for violating the  $n$-locality inequality can be written as
	\begin{equation}
		\label{unsharpmn}
		\lambda^{m}_{n,k}\geq \dfrac{m\lambda^{m}_{n,k-1}}{1+(m-1)\sqrt{1-\left( \lambda^{m}_{n,k-1}\right)^{2}}}
	\end{equation}  	
	From Eq.(\ref{nlocal}) we can find the  critical value of unsharpness parameter for the first observer as $\left(\lambda^{m}_{n,1}\right)^{\ast}= \left( \dfrac{\alpha_{m}}{2^{m-1}\sqrt{m}}\right)^n$, which is very small for large $n$.
	For every $m$, the critical value of the unsharpness parameter of the $(k-1)^{th}$ sequential edge observer we observe that $1+(m-1)\sqrt{1-\left( \lambda^{m}_{n,k-1}\right)^2} \geq m\sqrt{1-\lambda^{2}_{n,k-1}}$.  Using this in Eq.(\ref{unsharpmn}), one can demand that the lower bound on $\lambda_{n,k}$ requires
	\begin{equation}
		\label{unsharp}
		\lambda^{m}_{n,k}\geq \frac{\lambda^{m}_{n,k-1}}{\sqrt{1-\left(\lambda^{m}_{n,k-1}\right)^2}} 
	\end{equation}
	Note that Eq.(\ref{unsharp}) is obtained by approximation and, for small $n$, it overestimates the lower bound of the unsharpness parameter.
	Using the critical value $\left(\lambda^{m}_{n,1}\right)^{\ast}$ of Alice$_{1}^{1}$ and Eq.(\ref{unsharp}) we upper bound  the unsharpness parameter of Alice$^{1}_{2}$ for arbitrary $m$ input scenario as
	\begin{eqnarray}
		\label{44}
		\lambda^{m}_{n,2}&&\geq \frac{\left(\lambda^{m}_{n,1}\right)^{\ast}}{\sqrt{1-\left(\left(\lambda^{m}_{n,1}\right)^{\ast}\right)^2}}\\
		\nonumber
	\end{eqnarray}
	Next, using the Eq. (\ref{44}) in Eq.(\ref{unsharp}) for $k=3$ we can write
	\begin{eqnarray}
		\lambda^{m}_{n,3}&&\geq \frac{\left(\lambda^{m}_{n,2}\right)}{\sqrt{1-\left(\lambda^{m}_{n,2}\right)^2}} \geq\frac{1}{\sqrt{1/\left(\left(\lambda^{m}_{n,1}\right)^{\ast}\right)^{2}-2}}
	\end{eqnarray}
	By repeating this procedure $k^{th}$ times we obtain 
	\begin{eqnarray}
		\label{mnk}
		\lambda^{m}_{n,k} \geq\frac{1}{\sqrt{1/\left(\left(\lambda^{m}_{n,1}\right)^{\ast}\right)^{2}-k+1}}
	\end{eqnarray}
	Putting the critical value  $\left(\lambda^{m}_{n,1}\right)^{\ast}$ and performing a little manipulation of Eq. (\ref{mnk}) we have 
	\begin{eqnarray}
		\label{mnk1}
		\left( \dfrac{2^{m-1}\sqrt{m}}{\alpha_{m}}\right)^{2n}\geq \frac{1}{\left(\lambda^{m}_{n,k}\right)^{2}} +k-1
	\end{eqnarray}
	If the  $k^{th}$ sequential edge party's measurement is sharp, i.e., $\lambda^{m}_{n,k}=1$, we have  
	
	\begin{eqnarray}
		\label{mnk1}
		\left( \dfrac{2^{m-1}\sqrt{m}}{\alpha_{m}}\right)^{2n}\geq  k
	\end{eqnarray}Thus, for a given $k$, one can find a $n \equiv n(m,k)$ for which the nonlocality in star-network can be shared, so that 
	\begin{eqnarray}
		n(m,k)\geq\frac{\log_2 k}{2\log_{2}\left( \dfrac{2^{m-1}\sqrt{m}}{\alpha_{m}}\right)}
	\end{eqnarray}
	Note that for $m=2$, we recover the results in Eq. (\ref{nk}). Hence, given an arbitrary $k$ there exists a $n(m,k)$ for which the nonlocality can be shared by unbounded number of sequential observers across the one edge of the star-network.
	The quantity $\left(2^{m-1}\sqrt{m}\right)/\alpha_{m}>1$ for any arbitrary $m$ and saturates to $5/4$ for sufficiently large value of $m$. This implies that for large value of $m$ one has to take suitable large value of $n(m,k)$ to demonstrate the  sharing of nonlocality by unbounded number of sequential observers across one edge.   
	
	
	\section{Summary and discussions}
	\label{secVII}
	
	In this work, we demonstrated the sequential sharing of nonlocality in a quantum network through the quantum violation of bilocality and $n$-locality inequalities for two input and an arbitrary $m$ input scenarios.   We first considered the bilocality scenario that features two independent sources, two edge parties, and a central party. As an example, we provided an explicit derivation of optimal quantum violation of bilocality inequality by introducing an elegant SOS approach which enables us to fix the state and observables required for optimization. Note that no dimension is required to be assumed for this optimization. This approach can be extended for deriving the optimal quantum violation of  $n$-locality inequality for an arbitrary $m$ input scenario.
	
	We considered two different cases of sharing of nonlocality in the network; the symmetric case, when all the edge parties share the nonlocality, and the asymmetric case, when only one edge party shares the nonlocality. In the symmetric case, all the edge parties perform unsharp measurements sequentially, whereas in the asymmetric case, only one edge party is allowed to perform unsharp measurements, and the rest of the parties perform the sharp measurements. By first considering the two-input bilocality scenario, we demonstrated that the nonlocality can be shared by at most two sequential observers per edge party in the symmetric case. However, in the asymmetric case, when the sharing of only one edge party is considered, at most, six sequential observers can share nonlocality across one edge in the two-input ($m=2$) bilocality scenario ($n=2$). However, in the two-input  trilocality scenario ($n=3$) in star-network, we showed that nonlocality sharing in the asymmetric case can be extended to fourteen sequential observers across one edge party.
	
	We then considered the two-input $n$-locality scenario in a star-network configuration that features an arbitrary $n$ edge party and one central party. We demonstrated that in the symmetric case, the sharing of nonlocality in the $n$-locality scenario remains restricted to two observers for each edge party for any arbitrary $n$. However, in the asymmetric case, an unbounded number of sequential observers across one edge can share the nonlocality for a sufficiently large value of $n$.
	
	Further, we extended our study for the $n$-locality scenario in star-network for an arbitrary $m$ input scenario where each edge party receives an arbitrary $m$ input, and the central party receives $2^{m-1}$ inputs. We first showed for the three-input scenario ($m=3$) only one observer in each edge can share nonlocality in the symmetric case, and it is argued that for any $m\geq 3$ not more than one observer in each edge can share nonlocality in $n$-locality scenario. However, in the asymmetric case of sharing, an unbounded number of sequential observers can share the nonlocality across one edge for a sufficiently large value of $n$. In such a case, increasing the input $m$, one has to suitably increase the number of edge parties $n$ to demonstrate sharing an unbounded number of sequential observers. 
	
	We remark that there can be many different cases of sharing are possible. One may consider the sharing of nonlocality for two or more edges instead of the sharing across one edge in the asymmetric case. Following the general approach developed in this work, it will be straightforward to find the number of sequential observers who can share nonlocality in those cases.   
	
	
\section*{Acknowledgments}
S.S.M. acknowledges the UGC fellowship [Fellowship No.16-9(June 2018)/2019(NET/CSIR)]. AKP acknowledges the support from the project DST/ICPS/QuST/Theme 1/2019/4.

\end{document}